\documentclass[sigconf,anonymous=false]{acmart}
\pdfoutput=1

\usepackage[ruled,linesnumbered]{algorithm2e}
\usepackage[capitalise]{cleveref}
\usepackage{tabularx}
\usepackage{footnote}
\usepackage[flushleft]{threeparttable}
\usepackage{enumitem}
\usepackage{multirow}
\usepackage{graphicx,subfigure}

\graphicspath{ {./resources/} }

\newcolumntype{R}[1]{>{\raggedleft\arraybackslash}p{#1}}
\newcolumntype{L}[1]{>{\raggedright\arraybackslash}p{#1}}
\newcolumntype{C}[1]{>{\centering\arraybackslash}m{#1}}

\theoremstyle{definition}
\newtheorem{secDefinition}{Definition}[section]

\begin{document}

\title{MGTCOM: Community Detection in  Multimodal Graphs}

\author{Egor Dmitriev}
\email{e.dmitriev@students.uu.nl}
\affiliation{%
  \institution{Utrecht University}
  \city{Utrecht}
  \country{The Netherlands}
}
\author{Mel Chekol}
\email{m.w.chekol@uu.nl}
\affiliation{%
  \institution{Utrecht University}
  \city{Utrecht}
  \country{The Netherlands}
}
\author{Shihan Wang}
\email{s.wang2@uu.nl}
\affiliation{%
  \institution{Utrecht University}
  \city{Utrecht}
  \country{The Netherlands}
}

\begin{abstract}

Community detection is the task of discovering groups of nodes sharing similar patterns within a network.
With recent advancements in deep learning, methods utilizing graph representation learning and deep clustering have shown great results in community detection.
However, these methods often rely on the topology of networks (i) ignoring important features such as network heterogeneity, temporality, multimodality and other possibly relevant features.
Besides, (ii) the number of communities is not known a priori and is often left to model selection.
In addition, (iii) in multimodal networks all nodes are assumed to be symmetrical in their features; while true for homogeneous networks, most of the real-world networks are heterogeneous where feature availability often varies.
In this paper, we propose a novel framework (named MGTCOM) that overcomes the above challenges (i)--(iii). MGTCOM identifies communities through multimodal feature learning by leveraging a new sampling technique for unsupervised learning of temporal embeddings. Importantly, MGTCOM is an end-to-end framework optimizing network embeddings, communities, and the number of communities in tandem.
In order to assess its performance, we carried out an extensive evaluation on a number of multimodal networks. We found out that our method is competitive against state-of-the-art and performs well in inductive inference. 

\end{abstract}

\maketitle
\pagestyle{plain}


\section{Introduction}

Various systems can be modelled as complex networks such as 
social \cite{hagenCrisisCommunicationsAge2018}, citation \cite{ senCollectiveClassificationNetwork2008}, biological \cite{gosakNetworkScienceBiological2018} and transaction \cite{prykeAnalysingConstructionProject2004} networks.
The task of identifying patterns of nodes with common properties, in such networks, is known
as community detection.
There is an abundant number of community detection methods in literature that approach this problem through modularity optimization \cite{ schuetzMultistepGreedyAlgorithm2008}, clique identification \cite{ kumpulaSequentialAlgorithmFast2008}, and spectral optimization \cite{heLaplacianRegularizedGaussian2011}.
With recent advancements in graph representation learning a new type of methods have emerged which 
utilize context-based learning techniques (e.g., DeepWalk \cite{perozziDeepWalkOnlineLearning2014} or  Node2Vec \cite{groverNode2vecScalableFeature2016}) to obtain topology-aware node embeddings.
These embeddings are either combined with existing clustering methods \cite{ xueCrossdomainNetworkRepresentations2019}
or are jointly optimized with found clusters 
\cite{jiaCommunityGANCommunityDetection2019}
to obtain communities.

In the above studies, the multimodal characteristics of real-world networks are overlooked. 
These characteristics can manifest as meta-topological features (node and relation types) \cite{caoKnowledgePreservingIncrementalSocial2021}, temporal features, and contentual features (e.g., text and image attributes).
Introduction of multimodality contrasts \textit{homophily} assumed by previous methods as \textit{heterophily} and can play an essential role in detecting communities in multimodal networks, as connected nodes may belong to different communities when multiple feature types are considered \cite{zhuHomophilyGraphNeural}.  
While it is common for causal links to be present between these features, it cannot be assumed without extensive domain knowledge.
Various algorithms have been devised to address the issue of temporality and multimodality \cite{ luoDetectingCommunitiesHeterogeneous2021, liCommunityDetectionAttributed2018, faniUserCommunityDetection2020}, 
though as far as we are aware none of the methods are able to address the lossless setting where all the features are incorporated.

Another challenge is information variance present in heterogeneous real-world networks. 
Different node or relation types may have different feature subsets and/or dimensionality.
For instance, consider the Twitter dataset (SDS) we use in our experiments.
This network consists of users, tweets, hashtags, and various relations. 
Tweets have content as textual features and post dates as temporal features, while users only have biography as textual features, and hashtags have neither. 
Similarly, users can form a directed follower relation link, while multiple relations may be present between tweets such as retweet, mention or quote.
This shows that real-world networks can be incomplete and the goal is to devise a model that works under such circumstances. 

Overall, we would like to address some of the open problems in community detection (as pointed out in~\cite{suComprehensiveSurveyCommunity2021}). In particular, detecting communities when: networks are multimodal, some nodes have incomplete features (incompleteness constraints), the number of communities is not known a priori,   and learning node embeddings as well as finding communities in an end-to-end manner.   
Hence, we propose a novel community detection framework (MGTCOM) that is able to address these problems.
MGTCOM discovers communities through multimodal feature learning and unsupervised learning with a new temporal sampling technique. 
In particular, our key contributions include:
\begin{itemize}
    \item[(i)] A robust method for unsupervised inductive representation learning on multimodal networks.
    \item[(ii)] A new sampling technique, namely ballroom walk, for unsupervised learning of temporal embeddings.
    \item[(iii)] An end-to-end framework optimizing network embeddings, communities, and number of communities in tandem.
    \item[(iv)] Extensive evaluation on the quality of various features in multimodal networks.
    \item[(v)] We compare MGTCOM with state-of-the-art methods and demonstrate its robustness on inference tasks.
\end{itemize}

\section{Related Work} \label{sec:related_work}
A comparison of MGTCOM with the state-of-the-art is given in~\cref{tab:comparison_related_work}. These methods are chosen based on their popularity within the literature, recency and relevance to our task. From these methods we select several comparable ones as the baseline in our experiments.
As can be seen, MGTCOM is able to generate: (i) node, (ii) meta-topology, (iii) content, and (iv) temporal embeddings as well as (v) is able to infer the number of communities $(K)$. By contrast, state-of-the-art methods (such as GraphSAGE and ComE) are able to produce either two or three of the above. 
A commonality of all the methods is that they  all utilize topological features. 
Similarly, we focus on representation based community detection methods in contrast to traditional link-based methods.
We provide a brief introduction of these methods below. We refer the reader to the survey~\cite{suComprehensiveSurveyCommunity2021} for a detailed discussion.  
Throughout the paper we will use the terms network and graph interchangeably.

\begin{table}[t]
\centering
\caption{
    A comparison of MGTCOM with state-of-the-art. 
    \label{tab:comparison_related_work}
}
\resizebox{\columnwidth}{!}{
    \begin{tabular}{c|C{1.5cm}|C{1.5cm}|C{1.5cm}|C{1.5cm}|C{1.5cm}}
        \hline
        & topology & meta-topology    & content  & temporal   & infers $K$  \\ \hline
        GraphSAGE \cite{hamiltonInductiveRepresentationLearning2017}
        & $\bullet$   &                     & $\bullet$      &                  &           \\
        SageDy \cite{wuSageDyNovelSampling2021}
        & $\bullet$   &                     & $\bullet$      & $\bullet$        &           \\
        CTDNE \cite{nguyenContinuousTimeDynamicNetwork2018}
        & $\bullet$   &                     &                & $\bullet$        &           \\
        Metapath2Vec \cite{dongMetapath2vecScalableRepresentation2017} 
        & $\bullet$   &      $\bullet$               &                &         &           \\
        HGT \cite{huHeterogeneousGraphTransformer2020}
        & $\bullet$   & $\bullet$           & $\bullet$      &                  &           \\ \hline\hline
        ComE \cite{cavallariLearningCommunityEmbedding2017}
        & $\bullet$   &                     &                &                  &           \\
        GEMSEC \cite{rozemberczkiGEMSECGraphEmbedding2019}
        & $\bullet$   &                     &                &                  &           \\
        GRACE \cite{yangGraphClusteringDynamic2017}
        & $\bullet$   &                     & $\bullet$      &                  &           \\
        Fani et al. \cite{faniUserCommunityDetection2020}
        & $\bullet$   &                     & $\bullet$      & $\bullet$        &           \\
        CP-GNN \cite{luoDetectingCommunitiesHeterogeneous2021}
        & $\bullet$   & $\bullet$           &                &                  &           \\
        MGTCOM
        & $\bullet$   & $\bullet$           & $\bullet$      & $\bullet$        & $\bullet$ \\\hline
    \end{tabular}
}
\vspace{-4.5mm}
\end{table}

\smallskip
\textbf{Graph Embedding}.
%
Graph representation learning approaches such as DeepWalk 
and Node2Vec
utilize random walks as a means to generate context and adopt the Skip-gram \cite{mikolovDistributedRepresentationsWords2013} model to directly learn the node embeddings methods. 
By defining a trade-off between first- and higher-order proximity they provide a way to fine-tune the learned topological representations for the task at hand. 
Depth-first search sampling strategies (higher-order proximity) encourage network communities while breadth-first search (first-order proximity) encourages structural similarity as the local neighborhood is more thoroughly explored~\cite{groverNode2vecScalableFeature2016}. 

The above methods mainly work on homogeneous networks in which all nodes and edges belong to the same types. 
Heterogeneous networks contain \textit{meta-topological} information that characterizes various relationships between different types of nodes/entities.
%
One way to address meta-topological features is by using meta-path constrained random walks to capture semantic and structural relations between different node/entity types. An example of such a model is Metapath2Vec~\cite{dongMetapath2vecScalableRepresentation2017}. 
Meta-path describes a sequence of entity and relation types. 
%
Other methods utilize a representation-based approach \cite{bordesTranslatingEmbeddingsModeling2013} to explicitly capture meta-topological features by defining relations as translations between different node types. 
This approach is further utilized in graph convolutional networks (GCN)-based methods, such as GraphSAGE~\cite{hamiltonInductiveRepresentationLearning2017} and HGT~\cite{huHeterogeneousGraphTransformer2020},  to apply them in an inductive setting.
Furthermore, HGT improves the neighborhood sampling algorithm by introducing a type-based budget for unbiased sampling.

Temporal graph embedding approaches are mainly split into two categories. 
Snapshot-based approaches operate by temporally splitting the graph into multiple snapshots or subgraphs and applying (modifying) existing graph embedding methods by temporally smoothing between the snapshots \cite{parejaEvolveGCNEvolvingGraph2020}.
The second category are the continuous temporal representation approaches which attempt to capture temporal information within the learned embeddings.
Generally, these methods look at the temporal progression of individual nodes rather than utilizing predefined snapshots.
The techniques vary; for instance CTDNE introduces biased temporal random walks \cite{nguyenContinuousTimeDynamicNetwork2018}; and SageDy introduces a neighborhood sampling technique to filter for temporal neighborhood \cite{wuSageDyNovelSampling2021}.

\smallskip
\textbf{Community Detection}.
There is an abundance of work concerning the finding of community structures by relying mainly on topological features \cite{fortunatoCommunityDetectionGraphs2010}.
%
Recent community detection methods focus on exploiting feature-rich information found in multimodal networks \cite{suComprehensiveSurveyCommunity2021}. 
The focus has shifted from link-based methods towards deep learning methods which combine graph embedding methods with clustering algorithms (such as k-means or spectral clustering) \cite{tianLearningDeepRepresentations2014, kozdobaCommunityDetectionMeasure2015}. 
For instance, ComE~\cite{cavallariLearningCommunityEmbedding2017} uses a Gaussian mixture model to learn homogeneous graph embeddings and cluster parameters jointly while utilizing random walk based context sampling.
GEMSEC~\cite{rozemberczkiGEMSECGraphEmbedding2019} uses random walks to learn community structure and embeddings simultaneously on homogeneous graphs.
With the emergence of multimodal community detection methods heterophily is  important as similarity may not always be correlated with topological features (homophily) \cite{zhuHomophilyGraphNeural}.
In line with this, various methods \cite{liuCommunityDetectionBased2014, sunRelationStrengthAwareClustering2012} modify the Latent Dirichlet Allocation algorithm to incorporate attribute, topological and meta-topological information.
Cao et al. \cite{caoIncorporatingNetworkStructure2018} and Yang et al. \cite{yangGraphClusteringDynamic2017} utilize autoencoders to jointly optimize graph embeddings on content and topological information.
Fani et al. \cite{faniUserCommunityDetection2020} use topic models to construct a user interest histogram over a time axis, which in turn is used to learn temporal content-based node representations. 
These representations are interpolated with topological representations, the similarity along edges is computed, and fed as edge weight to the existing link-based community detection algorithm (namely Louvain~\cite{blondelFastUnfoldingCommunities2008}).
Meta-topological information can be used in various ways to assist in community detection, for instance, 
Luo et al. \cite{luoDetectingCommunitiesHeterogeneous2021} propose CP-GNN which combines a heterogeneous graph transformer architecture with k-means clustering to find communities in  heterogeneous graphs.

\smallskip
\textbf{Clustering}.
 Bayesian non-parametric methods such as Dirichlet Process Mixture (DPM) have had great results in clustering \cite{minSurveyClusteringDeep2018} and community detection tasks where the number of communities is unknown \cite{tonellatoBayesianNonparametricClustering2020}. 
As these models can evaluate the likelihood of a set of cluster parameters being drawn from a prior distribution, the task is transformed into a Markov Chain Monte Carlo  sampling problem. Since there are prohibitively many possible parameter states, various hierarchical algorithms are proposed to explore the most promising states efficiently \cite{ changParallelSamplingDP2013a}.
Because these methods can be estimated using Expectation-Maximization (EM) algorithms, the previously introduced embedding methods can be utilized to learn representations and clusters in an end-to-end manner \cite{cavallariLearningCommunityEmbedding2017, ronenDeepDPMDeepClustering2022}. MGTCOM takes advantage of this.

\section{Preliminaries} \label{sec:preliminaries}
\begin{figure*}[ht!]
\centering
\includegraphics[width=0.9\textwidth]{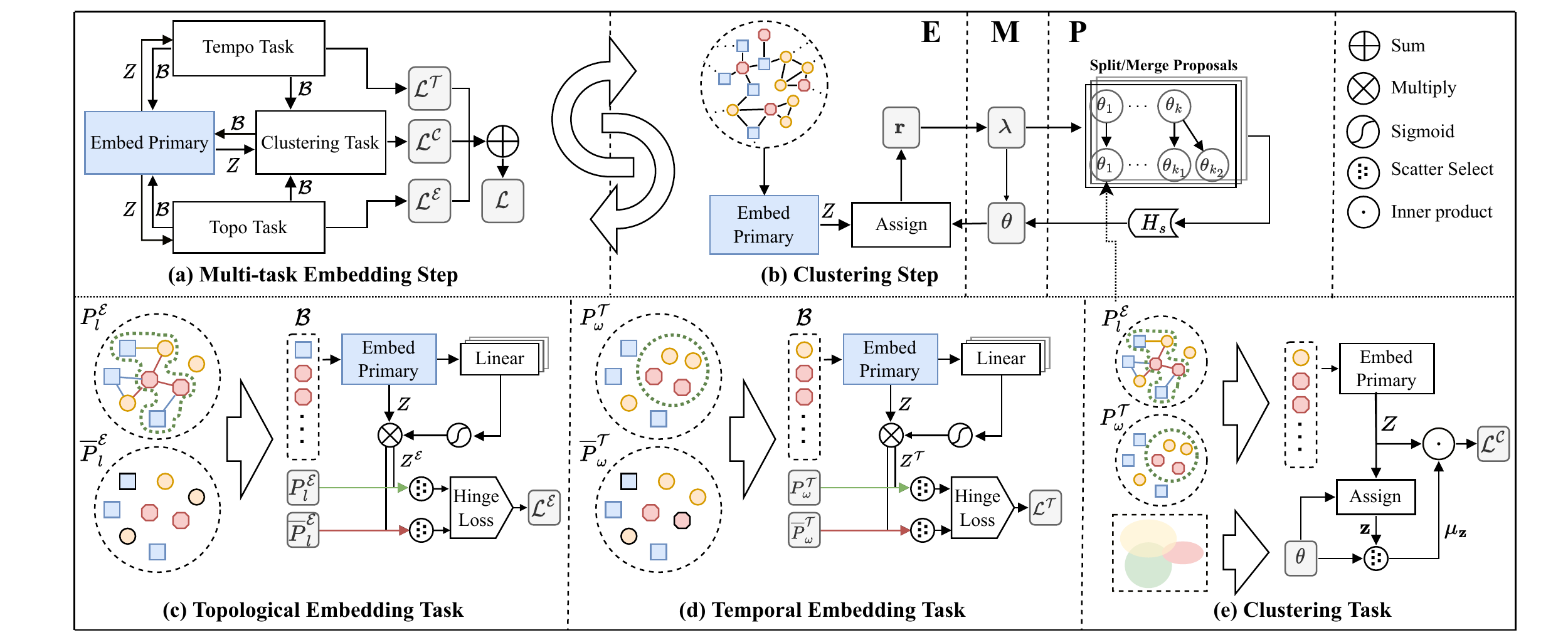}
\caption{Overview of the MGTCOM framework.
}
\label{fig:framework}
\end{figure*}

\begin{secDefinition}[\textbf{Heterogeneous graph}]
A heterogeneous graph, denoted by $G = (\mathcal{V}, \mathcal{E}, \mathcal{A}, \mathcal{R})$, consists of a set of nodes $\mathcal{V}$, a set of edges $\mathcal{E}$, and their associated type mapping functions $\phi : \mathcal{V} \rightarrow \mathcal{A}$ and $\psi : \mathcal{E} \rightarrow \mathcal{R}$. $\phi$ (resp. $\psi$) maps a node (resp. edge) to its type. $\mathcal{A}$ and $\mathcal{R}$ denote  sets of node and edge types, respectively, where $|\mathcal{A}| > 1$ and $|\mathcal{R}| > 1$. $G$ is a homogeneous if $|\mathcal{A}| = 1$ and $|\mathcal{R}| = 1$.
\end{secDefinition}

\noindent We define a multimodal information network by combining the notion of heterogeneous, continuous-time, and contentual networks. 
\begin{secDefinition}[\textbf{Multimodal graph}]
A multimodal graph is defined as $G = (\mathcal{V}, \mathcal{E}, \mathcal{A}, \mathcal{R}, \mathcal{T}, \mathcal{X})$ where $\mathcal{T}$ is a set of timestamps $t$ and $\mathcal{X}$ is a set of type specific feature matrices $\mathcal{X}_{\phi(\cdot)}$. 
A feature matrix is created by extracting and stacking feature vectors for each node given pre-trained dense representations of their image, text, or categorical attributes. 
Each node $v \in \mathcal{V}$ (resp. edge $e \in \mathcal{E}$) has a time range $\tau(v) = [t_s, t_e]$  (resp. $\tau(e) = [t_s, t_e]$), indicating the time period on which it is considered valid, where $t_s, t_e \in \mathcal{T}$. 
Moreover, 
each node $v$ has an attribute vector $\mathbf{x} \in \mathcal{X}_{\phi(v)}$.

\end{secDefinition}

\begin{secDefinition}[\textbf{Incompleteness constraints}] \label{def:incompleteness_constraints}
Real-world multimodal networks can be noisy, incomplete, and may change over time. 
In order to represent this information we introduce additional indicator functions to denote whether a node has a time interval $\mathbf{1}_{\mathcal{T}} : \mathcal{V} \rightarrow \{0, 1\}$, has a feature vector $\mathbf{1}_{\mathcal{X}} : \mathcal{V} \rightarrow \{0, 1\}$, or is unseen during training $\mathbf{1}_{\mathcal{V}} : \mathcal{V} \rightarrow \{0, 1\}$. 
We refer to the noisiness, incompleteness, and temporality as incompleteness constraints. 
\end{secDefinition}

\begin{secDefinition}[\textbf{Context window}]
A context window connects nodes based on some predefined criteria. Two nodes are \textit{context neighbors} if they occur in the same context window.
We use two different kinds of context windows. 
The first is the \textit{topological context window}. It connects two nodes $v_i$ and $v_j$ if there exists a h-hop path ${p}^{\mathcal{E}}_k$ in graph $G$ through which they are connected.
The second is \textit{temporal context window} ${p}^{\mathcal{T}}_{\omega}$. 
It connects $v_i$ and $v_j$ if they occur within a given time window $\omega = [t_s, t_e]$.
Going forward we use ${P}^{\mathcal{E}}_k$ and ${P}^{\mathcal{T}}_{\omega}$ to denote a fixed size sample of all possible context windows. We use contexts for learning task based node embeddings. 
\end{secDefinition}

\begin{secDefinition}[\textbf{Heterogeneous Graph Transformer}]

Classical GCNs focus mainly on homogeneous graphs~\cite{hamiltonInductiveRepresentationLearning2017}. For heterogeneous graphs, 
Heterogeneous Graph Transformer (HGT) \cite{huHeterogeneousGraphTransformer2020} adopts the transformer architecture \cite{vaswaniAttentionAllYou2017} by calculating mutual attention based on representation and meta-types of source nodes, target nodes and relation information.
 We utilize HGT to generate the primary embedding of a multimodal graph (\cref{sec:primary_embed}).
\end{secDefinition}

\vspace{-1mm}
\noindent\textbf{Problem formulation.} 
Given a multimodal graph $G$, our goal is to learn a node embedding function $\zeta: G_{v} \rightarrow \mathbb{R}^{d}$ which given a $h$-hop neighborhood subgraph $G_{v}$ of node $v$ produces a $d$-dimensional embedding vector $Z_v$.
The objective is to minimize the distance between embedding $Z_v$ to other node embeddings, given that they are topological and/or temporal context neighbors of node $v$.
Taking into account incompleteness constraints (\cref{def:incompleteness_constraints}), $\zeta$ should work under any valuation of $(\mathbf{1}_{\mathcal{X}(v)}, \mathbf{1}_{\mathcal{T}(v)}, \mathbf{1}_{\mathcal{V}(v)})$.
We also aim to find community parameters $\mathbf{\theta} = \{\mathcal{N}(\mu_1, \Sigma_1), ..., \mathcal{N}(\mu_K, \Sigma_K)\}$ and node-to-community assignment $\mathbf{z} \in \{0, \ldots, K\}^{|\mathcal{V}|}$ such that their members have a low inter-proximity in contrast to other nodes. 
Finally, the found community count $K$ should approximate the ground truth number of communities.
\section{The Proposed Approach}\label{sec:approach}

We present our framework for Community Detection in Multimodal Graphs (MGTCOM) that learns multimodal representation vectors for graph nodes and detects communities in tandem.
We achieve this by leveraging heterogeneous graph transformers \cite{huHeterogeneousGraphTransformer2020} to learn a primary node embedding function $\zeta$. In order to handle the incompleteness constraints, we introduce an auxiliary embedding vector $\mathbf{E}$ for known (or seen) nodes with missing features.
Next, we learn task-specific node representation for topological and temporal information by combining primary embeddings with task-specific transformation/attention and context sampling. 
As we utilize random walks for topological context sampling, we introduce its analogue as an unbiased temporal window sampling algorithm for temporal context collection.
Finally, we adopt DPMM (Dirichlet process mixture model) for community detection and close the loop by introducing cluster-based loss to ensure the graph embeddings are \textit{community-aware}.
The architecture of MGTCOM is given in \cref{fig:framework}. In the embedding step primary embeddings are used in auxiliary tasks to construct the multi-objective loss (\cref{fig:framework}(a)). 
(b) The clustering step updates clustering by alternating between Expectation (\textbf{E}), Maximization (\textbf{M}), and Proposal (\textbf{P}) steps.
(c) In the topological (topo) embedding task, random walk sampling and feature-wise attention minimize inter-node proximity.
(d) In the temporal (tempo) embedding task, ballroom walk sampling and feature-wise attention minimize proximity between temporally related nodes.
(e) Clustering task adds community awareness to the embeddings by minimizing proximity between nodes within the same cluster.

\subsection{Primary embedding module}\label{sec:primary_embed}
The central component of our framework is responsible for inferring the primary representation vector $Z_v$ given a node $v \in \mathcal{V}$ in a graph $G$. 
The basic idea is to learn node representation from its h-hop heterogeneous neighborhood subgraph $G_v$ while handling edge cases introduced by the incompleteness constraints (\cref{def:incompleteness_constraints}). 
The learning starts by sampling a subgraph $G_{\mathcal{B}}$ given a batch of central nodes $\mathcal{B}$  using the \textit{budget sampling} algorithm~\cite{huHeterogeneousGraphTransformer2020}.
The budget sampling algorithm works by restricting sampled subgraph at each layer given a per node type limit.
In our case, we define this limit as multiple $|\mathcal{B}|$  to avoid re-tuning its value for each dataset. 
\subsubsection{\textbf{Incompleteness constraints}}
Once the graph is sampled we split the task of initial feature inference into three cases to handle the incompleteness constraints (\cref{eq:feature_inf}).
(i) If a feature vector ($\mathbf{x}_v$) is present, then it is linearly projected into the representation space.
(ii) If the node $v$ is in the training set while no feature vector is present, then its representation is drawn from the \textit{auxiliary embedding} matrix $\operatorname{E}$. To avoid over reliance on the embeddings in preference for feature vectors we apply dropout 
on the resulting representation.
(iii) Finally, if an unseen node without a feature vector is encountered, the zero vector (denoted as $0_d$) is used, indicating that its feature vector has zero weight during the aggregation step of the graph convolution.
Note that for large datasets, it may not be feasible to keep a full auxiliary embedding matrix in memory. 
%
\par\nobreak 
\vspace{-0.2cm}
{%
\small 
\begin{align}
    H_v^{(0)} &= \begin{cases}
        \operatorname{Linear}(\mathbf{x_v}) & \mathbf{1}_{\mathcal{X}}(v) = 1 \\
        \operatorname{Dropout}(\mathbf{E}_s)& \mathbf{1}_{\mathcal{V}}(v) = 1 \\
        0_d                                 & \text{otherwise}
    \end{cases} \label{eq:feature_inf} \\
    H^{(l)} &= \operatorname{GeLU}(\operatorname{HGT}(G_{\mathcal{B}}, H^{(l-1)})) \label{eq:prim_conv}
\end{align}
}%
Next, given the subgraph $G_\mathcal{B}$ and the initial representation vector $H^{(0)}$,  $L$ layers of HGT graph convolutions are applied (\cref{eq:prim_conv}).
Each layer uses the representation vector of the previous layer and feeds its output through a $\operatorname{GeLU}$ \cite{hendrycksGaussianErrorLinear2020} activation function (chosen based on our experiments in which it outperforms other activation functions such as ReLU).
Finally,  the output vectors at the $L^\mathit{th}$ layer are used as primary embeddings for each node in the batch  $Z_{\mathcal{B}}$. 


\cref{alg:prim_feat} provides a full overview of the primary embedding algorithm. 
The basic idea is to learn node representation from its k-hop heterogeneous neighborhood subgraph $G_v$ while handling edge cases introduced by the incompleteness constraints (\cref{def:incompleteness_constraints}). 
The learning starts by sampling a subgraph $G_{\mathcal{B}}$ given a batch of central nodes using the \textit{budget sampling} algorithm on \cref{alg:pe:sampling}.
The \textit{budget sampling} algorithm works by restricting sampled subgraph at each layer given a per-node type limit.
For our use case, we define this limit as multiple $|\mathcal{B}|$ to avoid re-tuning its value for each dataset. 

Once the graph is sampled we split the task of initial feature inference into three cases to handle the incompleteness constraints.
(i) If a feature vector is present, then it is simply projected into the representation space.
(ii) If the node is in the training set while no feature vector is present, then its representation is drawn from the \textit{auxiliary embedding} matrix $\mathbf{E}$. To avoid overreliance on the embeddings in preference for feature vectors we apply dropout on the resulting representation.
(iii) Finally, if an unseen node without a feature vector is encountered, the zero vector (denoted as $0_d$) is used, indicating that its feature vector has zero weight during the aggregation step of the graph convolution.

\begin{algorithm}[!t]
    \small
    \caption{Batchwise primary node embedding}\label{alg:prim_feat}
    \SetKwFunction{procPrimary}{EmbedPrimary}

    \SetKwProg{myproc}{Procedure}{}{}
    \myproc{\procPrimary{}}{
        \KwIn {
            multimodal graph $G = (\mathcal{V}, \mathcal{E}, \mathcal{A}, \mathcal{R}, \mathcal{X})$,
            mini-batch $\mathcal{B} \subseteq \mathcal{V}$,
            auxiliary node embedding $\mathbf{E} \in \mathbb{R}^{N_{\bar{\mathcal{X}}} \times d }$
            where $N_{\bar{\mathcal{X}}} = |\{v|v \in \mathcal{V}, \mathbf{1}_{\mathcal{X}}(v) = 0\}|$,
            number of convolutional layers $L$
        }
        \KwOut {
            The primary embedding $Z_{\mathcal{B}}$ for nodes in batch $\mathcal{B}$
        }
            $G_{\mathcal{B}}(\mathcal{V}_{\mathcal{B}}, \mathcal{E}_{\mathcal{B}}, \mathcal{A}_{\mathcal{B}}, \mathcal{R}_{\mathcal{B}}, \mathcal{X}_{\mathcal{B}}) \leftarrow \operatorname{HeteroSample}(G, \mathcal{B}, L)$\; \label{alg:pe:sampling}
            \For {$s \in \mathcal{V}_{\mathcal{B}}$} {
                $H_s^{(0)} = \begin{cases}
                    \operatorname{Linear}(\mathbf{x_s}) & \mathbf{1}_{\mathcal{X}}(s) = 1 \\
                    \operatorname{Dropout}(\mathbf{E}_s)& \mathbf{1}_{\mathcal{V}}(s) = 1 \\
                    0_d                                 & \text{otherwise}
                \end{cases}$\; \label{alg:pe:feature_inf}
            }
            \For {$l = 1$ \KwTo $L$} {
                $H^{(l)} = \operatorname{GeLU}(\operatorname{HeteroConv}(G_{\mathcal{B}}, H^{(l-1)}))$\; \label{alg:pe:conv}
            }
            $Z_{\mathcal{B}} = \{H^{(L)}_t | t \in \mathcal{B} \}$\; \label{alg:pe:batch}
        \Return {$Z_{\mathcal{B}}$}
    }
\end{algorithm}

On \cref{alg:pe:conv}, given the subgraph $G_\mathcal{B}$ and the initial representation vector $H^{(0)}$,  $L$ layers of HGT graph convolutions are applied.
Each layer uses the representation vector of the previous layer and feeds its output through a $\operatorname{GeLU}$ \cite{hendrycksGaussianErrorLinear2020} activation function.
Finally, the output vectors at the $L^\mathit{th}$ layer are used as primary representation vectors and output for each query node in the batch on \cref{alg:pe:batch}. 

\subsection{Multi-task representation learning}
During task-specific learning we focus on two main tasks capturing the intricacies of multimodal networks.
The topological task denoted by $\mathcal{L}^\mathcal{E}$ focuses on minimizing the representation distance between nodes that are proximate within the network (\cref{fig:framework}(b)).
Analogously, the temporal task $\mathcal{L}^\mathcal{T}$ focuses on minimizing the distance between nodes that co-occur at the same time range (\cref{fig:framework}(b)).
While fundamentally different since the tasks are trained in parallel, they benefit from weight sharing and from node sharing during primary embedding as the subgraph batches are centered around the same nodes.

\subsubsection{Task-based attention}
An important observation is that while temporal and topological communities are both important during analysis, they are not always correlated.
In most of the benchmarking datasets such as  DBLP temporal features and topology show low correlation. However, it is very rare that contentual features are completely independent of topology and temporality.
Given this observation, we admit that it may not be possible to train a model that excels at both tasks.
To work around this issue while still capturing both tasks in a single embedding vector we introduce \textit{task-based attention}.
The basic idea is that while primary embedding extracts suitable features from the multimodal network, task-specific attention selects the most relevant of these features for the task at hand.
Inspired by transformers \cite{vaswaniAttentionAllYou2017} we define multi-head attention $\operatorname{\textit{att}}^{task}$ to capture various feature patterns \cref{eq:task_attention}.
The task-based transformation function  is defined as \cref{eq:task_transform} where the primary representation vector is attended to using matrix multiplication.
We specialize this function for topological task as $f^{\mathcal{E}}(\mathbf{Z})$ producing $\mathbf{Z}^{\mathcal{E}}$ and temporal task $f^{\mathcal{T}}(\mathbf{Z})$ producing $\mathbf{Z}^{\mathcal{T}}$.
\par\nobreak 
\vspace{-0.25cm}
{
\small 
\begin{align}
    f^{task}(\mathbf{Z}) &= \mathbf{Z} \cdot \operatorname{\textit{att}}^{task}(\mathbf{Z}) \label{eq:task_transform}  \\
    \operatorname{\textit{att}}^{task}(\mathbf{Z}) &= \underset{i \in [1..h]}{\|} \sigma \left[ \operatorname{Linear}_i^{task}(\mathbf{Z}) \right] \label{eq:task_attention}
\end{align}
}%
where $\|$ denotes concatenation operation, $h$ is the number of attention heads, and $\sigma$ denotes nonlinearity (sigmoid function). 
\subsubsection{Objective function} \label{sec:obj_fn}
In order to learn model parameters in an unsupervised way, we define contrastive loss.
Task-specific positive context sample $P$ and a negative context sample $\bar{P}$, both sharing a central query node $q$ are used to construct positive $(q, p)$ and negative $(q, n)$ node pairs respectively.
We define an objective function based on max-margin (Hinge loss)  (\cref{eq:max_margin}) which aims to maximize the inner product similarity between the query and positive examples.
On the other hand, the inner product of query and negative samples is minimized to be smaller than that of the positive samples by some predefined value of $\Delta$ (we performed hyperparameter analysis to find the best).
To smoothen out the noise caused by context sampling, we average similarity over negative samples within the max loop (\cref{eq:mm_mean_aff}).

\begin{align}
    \operatorname{MM-Loss}(\mathbf{Z}, P, \bar{P}, q) &= \max_{n \in \bar{P}} \left\{0, \mathbf{Z}_q \mathbf{Z}_n - \widetilde{\mathbf{Z}_q \mathbf{Z}_p} + \Delta \right\} \label{eq:max_margin} \\
    \widetilde{\mathbf{Z}_q \mathbf{Z}_p} &= \frac{1}{|P|} \sum_{p \in {P} } \left(\mathbf{Z}_q \mathbf{Z}_p \right) \label{eq:mm_mean_aff}
\end{align}


\begin{figure}[th!]
\centering
\subfigure[]{
\includegraphics[width=0.16\textwidth]{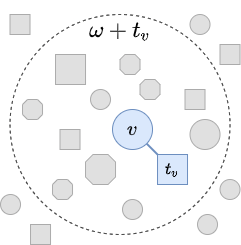}
}\quad
\subfigure[]{
\includegraphics[width=0.16\textwidth]{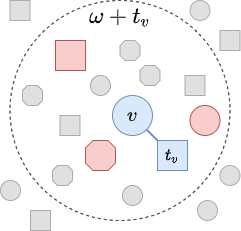}
}\\
\subfigure[]{
\includegraphics[width=0.16\textwidth]{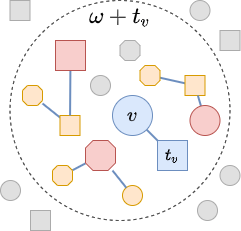}
}\quad
\subfigure[]{
\includegraphics[width=0.16\textwidth]{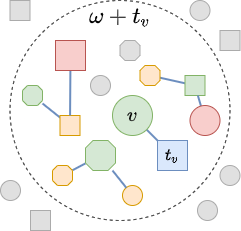}
}

\caption{
    Visual overview of the Ballroom Walk temporal sampling algorithm.
}
\label{fig:br}
\end{figure}
\begin{algorithm}[!t]
    \small
    \caption{Ballroom walk sampling}\label{alg:ballroom_walk}
    \SetKwFunction{algo}{BallroomWalk}
    \SetKwFunction{proc}{TemporalRW}
    \SetKw{KwGoTo}{go to}
    
    \SetKwProg{myalg}{Algorithm}{}{}
    \myalg{\algo{}}{
    \KwIn {
        multimodal graph $G = (\mathcal{V}, \mathcal{E}, \mathcal{A}, \mathcal{R}, \mathcal{X})$,
        relative temporal window $\omega$,
        walks per node $n$,
        walk length $l$,
        center node $v$, 
    }
    \KwOut {
        Temporal $n$ random walks $P_l$
    }
        Initialize $C$\;
        $t_v = \begin{cases}
            t_v \sim \tau(v)                                            & \mathbf{1}_{\mathcal{T}}(v) = 1 \\
            \proc{$v$, $(-\infty, \infty)$}     & \text{otherwise}
        \end{cases}$\; \label{alg:brw:infer_t}
        $N(v) = \{u|
            u \in \mathcal{V}, 
            \tau(u) \cap \omega + t_v \neq \emptyset
        \}$\; \label{alg:brw:neighborhood}
        \For {$i = 1$ \KwTo $n$}{
            $m \sim \mathcal{N}$\;
            $C = C \cup \proc{$m$, $\omega + v_t$}$\; \label{alg:brw:context}
        }
        \texttt{RandomPermute}($C$)\;

        \For {$i = 1$ \KwTo $n$}{
            $P_l = \{C_j | i \cdot l \leq j < i \cdot l+l\}$\; \label{alg:brw:walks}
            \KwRet $P_l$\;
        }
    }{}
    %
        



\end{algorithm}
 \vspace{-0.1cm}

\smallskip
\noindent
\textit{Topological context window sampler}. 
The objective of task-specific learning is mainly defined by the context (window) sampling method.
As our method allows for inference of primary and task-specific representations for unseen nodes, we assume that topological, meta-topological and contentual features contain enough information or are correlated with the objectives of the tasks.
%
To gather the topological context $P^{\mathcal{E}}$, Node2Vec biased random-walk algorithm is utilized \cite{groverNode2vecScalableFeature2016}.
By choosing a low value for its control parameter $q$ we discourage structural/topological equivalence representation in favor of larger neighborhood exploration (depth-first strategy) which is useful for community representation.
In our framework, the query nodes and negative contexts are shared across both tasks. In order to collect temporal context $P^{\mathcal{T}}$ of size $l$, we introduce ballroom walk sampling below.  

\subsubsection{Two temporal context samplers}
\label{sec:tempo_sampling}

Temporal features are often not correlated with network topology.
We propose a separate context sampling method that, given a query node and an interval window $\omega$, returns other nodes occurring within the same time window. 
$\omega$ is determined by using the dataset statistics as a fraction of the complete time range $\mathcal{T}$.
By picking a small enough interval window, a fine-grained continuous-time representation vector can be learned.
This is because additional granularity is achieved by centering the sample around the query node.
Edge cases arising from the \textit{incompleteness constraints} need to be handled where the nodes are missing timestamps $\mathbf{1}_{\mathcal{T}} = 1$.
While the usual semantic approach is to consider these nodes omnipresent (static), the naive window sampling methods quickly get congested with static to static context pairs.
Our aim is to alleviate this issue using biased sampling in favor of non-static pairs.

\smallskip
\noindent
\textbf{Temporal random walk context sampler}. We start by introducing the  temporal random walk procedure  which enforces standard random walks over the network to stay within a predetermined time window $\omega$.
Here random walk of size $l$ is constructed by picking a randomly connected node to the current head node  within a window.
If no such node is present, then the random walk is restarted from any already picked node.
The walk is extended with a new head node until it reaches the desired length.

\medskip
\noindent
\textbf{Ballroom walk context sampling}. Since temporal random walk looks naively for only adjacent temporal neighbors, we propose a more complex sampling method called \textit{ballroom walk} that searches for temporal neighbors which may not necessarily be connected.  
The outline of the proposed method is shown in \cref{alg:ballroom_walk}. 
It starts by inferring the sampling timestamp $t_v$  picking a random timestamp in which the center/query node $v$ occurs in.
If the node is static, the timestamp of its nearest neighbor reachable through temporal random walk ($\operatorname{TemporalRW()}$) is selected (\cref{alg:brw:infer_t}).
In order to reliably sample the temporal neighborhood, $n$ root nodes  are picked occurring in the time-window relative to the sampling timestamp (\cref{alg:brw:neighborhood}).
Temporal context $C$ is constructed by collecting temporal random walks starting from root nodes $m$ given a relative time window $\omega + t_v$ (\cref{alg:brw:context}). 
Finally, $l$ long context paths are created as random subsets of $C$. 
Note that because a sampled context is valid for all member nodes, random walk-like throughput optimization can be used by setting a larger window length than context size~\cite{perozziDeepWalkOnlineLearning2014}. An example of ballroom walk sampling is given in \cref{fig:br}. In \cref{fig:br}(a) the sampling timestamp $t_v$ for query node $v$ is inferred given the nearest neighbor if the node is static (blue). The relative time window is determined as $\omega + t_v$.
    (b) The root context nodes are sampled from the relative time window (red).
    (c) Context is extended with temporal random walks from the root nodes (yellow).
    (d) The context path is sampled from the collected context (green).

Due to timestamp inference, the first- and second-order proximity static to static pairs are ignored.
By only passively sampling omnipresent nodes we mitigate the over-saturation issue while still being fair.
Most importantly the neighborhood of central nodes is being sampled independently of their topology.
By sampling within a temporal window, we avoid not relying on the correlation of temporality with topology.

\subsection{Community detection with unknown $K$}
For community detection, we adopt the DPMM split/merge algorithm proposed by Chang and Fisher III \cite{changParallelSamplingDP2013a}.
In our implementation, we use Normal Wishart (NW) as a conjugate prior and use variational lower bound in our convergence criteria.
Specifically, we monitor the log sum of the variational lower bound 
for the supercluster and subcluster models. 
The variational lower bound is computed as the product of variational distribution $q(\mathbf{z})$, the normalizing constant of the Dirichlet distribution $B(\alpha_0)$, and the normalizing constant of the Normal Wishart distribution $C(W, \nu)$.
Once its monitored value starts oscillating, then the model has converged and is moved into the proposal state.
If the model parameters remain unchanged during the proposal stage (no split or merge is accepted), then the clustering is complete.
The only parameters relevant for our clustering method are the prior hyperparameters. 
Most of the other parameters (i.e. $\alpha$, $\kappa$, and $\nu$) are not very relevant if they are much smaller than the sample count.
We use the $\Sigma_{scale}$ parameter to scale the dataset covariance for more effective control over the strength of the data-bound prior parameters $\mu_0$ and $\Sigma_0$.
%
To fit the clustering model we use primary embedding to calculate assignment and posterior parameters as it contains features relevant for both temporal and topological tasks.
It is worth noting that it is not necessary to have all the embeddings in memory as exact posterior parameters depend on data $\mu$ and $\Sigma$ which can be calculated over multiple batches.

\subsection{End-to-end framework}
Given a graph embedding, it is straightforward to find communities by performing the embedding and clustering tasks sequentially. 
This approach lacks a unified objective, thus, the node embeddings may not be optimized for community detection.
We extend the objective with cluster-based loss calculated as the distance between a node embedding and its assigned cluster $z_v$ \cref{eq:loss_clus}.
This introduces a feedback loop that encourages the model to reinforce community structures while optimizing the topological and temporal objectives \cref{eq:combined_loss}.
The influence of three objectives can be controlled using hyperparameters $\beta^{\mathcal{E}}$, $\beta^{\mathcal{T}}$, and $\beta^{\mathcal{C}}$.

\begin{align}
    \mathcal{L}^{C} &= \left\|Z_v - \mu_{z_v} \right\|^2_{\ell 2} \label{eq:loss_clus} \\
    \mathcal{L}^{\mathcal{E}} &= \operatorname{MM-Loss}(\mathbf{Z}, P^{\mathcal{E}}, \bar{P}, v) \label{eq:loss_topo} \\
    \mathcal{L}^{\mathcal{T}} &= \operatorname{MM-Loss}(\mathbf{Z}, P^{\mathcal{T}}, \bar{P}, v) \label{eq:loss_tempo}\\
    \mathcal{L} &= \beta^{\mathcal{E}} \mathcal{L}^{\mathcal{E}} + \beta^{\mathcal{T}} \mathcal{L}^{\mathcal{T}} + \beta^{\mathcal{C}} \mathcal{L}^{\mathcal{C}} \label{eq:combined_loss}
\end{align}

\noindent 
With this closed feedback loop, the training procedure consists of two alternating stages (See \cref{fig:framework}(a) and (b)). See \cref{alg:pipeline} for a detailed overview.
The \textit{embedding optimization} stage (\cref{alg:pipeline:feat_optimization}), is responsible for optimizing the graph embedding function parameters while keeping cluster parameters $\theta$ fixed (\cref{alg:pipeline:cluster_optimization}).
Once the graph embeddings are updated we run $I_c$ clustering/EM steps to optimize cluster parameters $\theta$ while keeping node representations fixed. Note that the \textit{representation optimization} stage is run until convergence as part of pretraining beforehand to ensure the clusters are initialized properly.

\begin{algorithm}[t!]
    \small
    \caption{MGTCOM learning pipeline}\label{alg:pipeline}
      \KwIn {
            multimodal graph $G = (\mathcal{V}, \mathcal{E}, \mathcal{A}, \mathcal{R}, \mathcal{X})$,
            number of convolutional layers $L$
        }
        \KwOut {
            Embedding $Z$ for nodes in  $G$ and clusters parameters $\theta$
        }
    \For {$subiter = 1$ \KwTo $I$}{
        \For {$v \in \mathcal{V}$}{ \label{alg:pipeline:feat_optimization}
            \tcc{Gather context samples}
            $P^{\mathcal{E}} = \texttt{Node2VecRandomWalk}(G, l, v)$\;
            $P^{\mathcal{T}} = \texttt{BallroomWalk}(G, \omega, l, v)$\; 
            $\overline{P} \stackrel l\sim \mathcal{V}$ Negative sampling\;
            $\mathcal{B} = P_l^{\mathcal{E}} \cup P_l^{\mathcal{T}} \cup \bar{P}_l$\;
            $\mathbf{Z} = \texttt{PrimaryEmbed}(G, \mathcal{B})$\;
            Compute task embeddings $Z^{\mathcal{E}}$, $Z^{\mathcal{T}}$ using \cref{eq:task_transform}\;
            Compute loss $\mathcal{L}^{\mathcal{E}}$, $\mathcal{L}^{\mathcal{T}}$, $\mathcal{L}^{\mathcal{C}}$, $\mathcal{L}$ using \cref{eq:loss_clus,eq:loss_topo,eq:loss_tempo,eq:combined_loss} given respective context $P^{\mathcal{E}}$, $P^{\mathcal{T}}$\;
        }
        \For {$iter = 1$ \KwTo $I_c$}{ \label{alg:pipeline:cluster_optimization}
            \If {$i = 1$}{
                Initialize $\theta$ using K-means
            }
            Update $\theta$ using EM given $Z$
        }
    }
\end{algorithm}
\section{Experiments} \label{sec:experiments}
\begin{table}[t!]
\centering
\caption{
Dataset statistics. 
}
\label{tab:datasets}
\small
\resizebox{\columnwidth}{!}{
\begin{tabular}{@{}ccccccc@{}}
    \toprule
    Dataset                 & Node type     & \# Nodes  & Edge type             & \# Edges  & Temporal                      & Labelled \\
    \midrule
    \multirow{3}{*}{DBLP}   & Author (A)    & 5,162     & A - Authored - P      & 11,022    & \multirow{3}{*}{$\bullet$}    & \multirow{3}{*}{$\bullet$}\\
                            & Paper (P)     & 5,511     & P - Published In - V  & 5,511     &                               &                           \\
                            & Venue (V)     & 14        &                       &           &                               &                           \\ \midrule
    \multirow{3}{*}{IMDB}   & Person (P)    & 8,491     & P - Directed - M      & 4,939     & \multirow{3}{*}{$\bullet$}    & \multirow{3}{*}{}         \\ 
                            & Movie (M)     & 5,043     & P - Acted In - M      & 15,086    &                               &                           \\
                            & Genre (G)     & 26        & M - Tagged   - G      & 14,504    &                               &                           \\ \midrule
    \multirow{7}{*}{SDS}    & User (U)      & 34,919    & U - Tweeted  - T      & 56,173    & \multirow{7}{*}{$\bullet$}    & \multirow{7}{*}{}         \\ 
                            & Hashtag (H)   & 2,341     & T - Reply To - U      & 21,769    &                               &                           \\
                            & Tweet (T)     & 56,173    & T - Reply To - T      & 4,296     &                               &                           \\
                            &               &           & T - Quote    - T      & 882       &                               &                           \\
                            &               &           & T - Mention  - U      & 70,367    &                               &                           \\
                            &               &           & T - Mention  - H      & 12,313    &                               &                           \\
                            &               &           & U - Follows  - U      & 5,649,098 &                               &                           \\ \midrule
    \multirow{1}{*}{ICEWS}  & Entity (E)    & 10,463    & 123 different types   & 915,028   & \multirow{1}{*}{$\bullet$}    & \multirow{1}{*}{}         \\ 
    \bottomrule
\end{tabular}
}
\vspace{-0.1cm}
\end{table}
\begin{table*}[t!]
\centering
\caption{
    Comparison of performance of baselines on multimodal graph learning tasks. 
    ("-" means no data available, for example metapaths are not available for ICEWS).
    The calculated metrics are the link prediction accuracy ($LP_{ACC}$), predictive accuracy on ground truth communities $CF_{ACC}$ $L_y$, timestamp predictive accuracy $CF_{ACC}$ $L_\mathcal{T}$, NMI score of detected communities ($COM_{NMI}$) given predefined communities ($L_y$, $L_\mathcal{T}$, $L_G$) and Modularity.
}
\label{tab:results_perf}
 \resizebox{0.86\textwidth}{!}{
\begin{tabular}{cccccccccccc}
\toprule
Dataset    &                        & GraphSAGE & Node2Vec      & Metapath2Vec  &   ComE        & GEMSEC    &  CTDNE    &  CP-GNN   & MGTCOM  & MGTCOM$^{\mathcal{T}}$ & MGTCOM$^{\mathcal{E}}$ \\
\midrule
\multirow{8}{*}{DBLP}
    & $LP_{ACC}$                    & 0.624     & 0.710         & 0.624         & 0.735         & 0.544     & 0.701     & 0.522     & 0.743         & 0.634  & \textbf{0.794} \\
    & $CF_{ACC}$ $L_y$              & 0.315     & 0.832         & 0.696         & 0.842         & 0.831     & 0.809     & 0.506     &\textbf{0.896} & 0.330  & 0.884 \\
    & $CF_{ACC}$ $L_\mathcal{T}$    & 0.309     & 0.308         & 0.296         & 0.328         & 0.324     & 0.488     & 0.313     &\textbf{0.758} & 0.508  & 0.320 \\
    & $COM_{NMI}$ $L_y$             & 0.051     &\textbf{0.549} & 0.508         & 0.463         & 0.385     & 0.537     & 0.209     & 0.465         & 0.059  & 0.492 \\
    & $COM_{NMI}$ $L_\mathcal{T}$   & 0.006     & 0.033         & 0.024         & 0.025         & 0.022     & 0.059     & 0.022     &\textbf{0.209} & 0.168  & 0.026 \\
    & $COM_{NMI}$ $L_G$             & 0.040     & 0.425         & 0.413         &\textbf{0.470} & 0.314     & 0.401     & 0.107     & 0.336         & 0.039  & 0.371 \\
    & Modularity                    & 0.028     &\textbf{0.662} & 0.583         & 0.636         & 0.492     & 0.642     &-0.035     & 0.427         & 0.137  & 0.514 \\ \midrule
\multirow{6}{*}{ICEWS}
    & $LP_{ACC}$                    & 0.525     & 0.936         &-              & 0.880         & 0.768     & 0.921     & 0.709     & 0.903         & 0.896  & \textbf{0.945}   \\
    & $CF_{ACC}$ $L_\mathcal{T}$    & 0.294     & 0.301         &-              & 0.264         & 0.310     & 0.285     & 0.273     & 0.316         &\textbf{0.318} & 0.313     \\
    & $COM_{NMI}$ $L_\mathcal{T}$   & 0.018     & 0.040         &-              & 0.015         & 0.022     & 0.022     & 0.013     &\textbf{0.057} & 0.002  & 0.011            \\
    & $COM_{NMI}$ $L_G$             & 0.227     & 0.354         &-              & \textbf{0.548}& 0.309     & 0.347     & 0.204     & 0.119         & 0.001  & 0.447            \\
    & Modularity                    & 0.218     & 0.215         &-              & \textbf{0.483}& 0.311     & 0.239     & 0.199     & 0.007         & 0.001  & 0.390            \\ \midrule
\multirow{6}{*}{IMDB}
    & $LP_{ACC}$                    & 0.714     & 0.757         & 0.642         & 0.666         & 0.637     & 0.728     & 0.598     & 0.721         & 0.724  & \textbf{0.773} \\
    & $CF_{ACC}$ $L_\mathcal{T}$    & 0.346     & 0.373         & 0.309         & 0.394         & 0.380     & 0.488     & 0.316     &\textbf{0.659} & 0.556  & 0.377 \\
    & $COM_{NMI}$ $L_\mathcal{T}$   & 0.022     & 0.025         & 0.008         & 0.031         & 0.013     & 0.065     & 0.004     &\textbf{0.239} & 0.231  & 0.026 \\
    & $COM_{NMI}$ $L_G$             & 0.039     & 0.181         & 0.038         &\textbf{0.197} & 0.094     & 0.160     & 0.033     & 0.107         & 0.031  & 0.158 \\
    & Modularity                    &-0.172     & 0.190         & 0.070         &\textbf{0.395} & 0.073     & 0.196     & 0.053     & 0.119         & 0.114  & 0.286 \\\midrule
\multirow{6}{*}{SDS}
    & $LP_{ACC}$                    & 0.922     & 0.953         & 0.853         & 0.758         & 0.878     & 0.955     &     -     & 0.934         & 0.616  & \textbf{0.956} \\
    & $CF_{ACC}$ $L_\mathcal{T}$    & 0.521     & 0.445         & 0.399         & 0.386         & 0.384     & 0.447     &     -     & 0.523         & \textbf{0.887}& 0.492 \\
    & $COM_{NMI}$ $L_\mathcal{T}$   & 0.250     & 0.149         & 0.018         & 0.117         & 0.015     & 0.161     &     -     & 0.204         & \textbf{0.536}  & 0.044 \\
    & $COM_{NMI}$ $L_G$             & 0.186     & 0.277         & 0.024         & 0.346         & 0.117     & 0.233     &     -     & 0.120         & 0.043  & \textbf{0.389} \\
    & Modularity                    & 0.088     & 0.163         & -0.003        & 0.301         & 0.037     & 0.016     &     -     & 0.015         & 0.005  & \textbf{0.374} \\
\bottomrule
\end{tabular}
}
\end{table*}

For the sake of space, we only report some key results and do not report our extensive ablation study. Our code repository is available at \href{https://github.com/EgorDm/MGTCOM}{https://github.com/EgorDm/MGTCOM} and all the datasets can be accessed at \href{https://zenodo.org/record/7273761}{https://zenodo.org/record/7273761}.

We investigate the effectiveness of MGTCOM by evaluating its performance on auxiliary tasks related to multimodal networks.
%
In order to measure predictive performance over distinct aspects of our data, we first identify the following labels, then describe the auxiliary tasks (\S\ref{sec:eval_metrics}). We have three labels: \textit{ground truth labels ($L_y$)}, \textit{node timestamps $L_\mathcal{T}$} (we split the nodes evenly into snapshot labels given the timestamp of their first occurrences), and  \textit{Link-based communities $L_G$}. For the $L_G$, we first identify community labels using the Louvain method~\cite{blondelFastUnfoldingCommunities2008}. Then we use those labels to assess the quality of individual node embeddings for community detection. As the Louvain method greedily approximates optimal communities, we don't use this label for formal comparison.

\subsection{Evaluation metrics}\label{sec:eval_metrics}
There are no measures that can assess the quality of communities in multimodal networks.
Therefore, we evaluate our model component-wise by defining related auxiliary tasks. 
On a high level, these tasks evaluate the efficiency of topological and temporal node embeddings and found communities. 

\begin{itemize}[leftmargin=*]

\item \textit{Classification (CF)}.
Given a set of node embeddings and their respective ground truth labels, we train a logistic regression model to predict node labels. 
For the predicted node labels, we calculate the average classification accuracy. 

\item \textit{Link prediction (LP)}.
Given a set of positive and negative node pairs, binary classification is used to predict whether an edge exists within the graph.
We use a held-out positive and randomly sampled negative sets of edges to train a logistic regression model.
The inner-product similarity between a pair of node embeddings is used as input for the model.
By repeating this process three times, the average accuracy is calculated.

\item\textit{Link-based community quality}.
We use modularity 
to measure the quality of topological communities~\cite{girvanCommunityStructureSocial2002}. 

\item\textit{Ground-truth community quality (COM)}.
To asses the quality of detected communities for specific tasks, we measure the Normalized Mutual Information (NMI) score given a task-based label.
\end{itemize}

\subsection{Experimental setup} \label{sec:exp_setup}
As shown in \cref{tab:comparison_related_work}, there are no directly comparable methods to ours in terms of features.
For a fair and coherent comparison, we define three variants of  MGTCOM. 
In addition to the complete end-to-end model MGTCOM, we split our framework into a temporal model MGTCOM$^{\mathcal{T}}$ and topological model MGTCOM$^{\mathcal{E}}$, by removing $\mathcal{L}_{\mathcal{T}}$ and $\mathcal{L}_{\mathcal{E}}$ from the objective respectively.

\begin{figure*}[t!]

\begin{minipage}{.67\textwidth}
    \subfigure[$MGTCOM$ + $L_y$]{
    \includegraphics[width=0.23\textwidth]{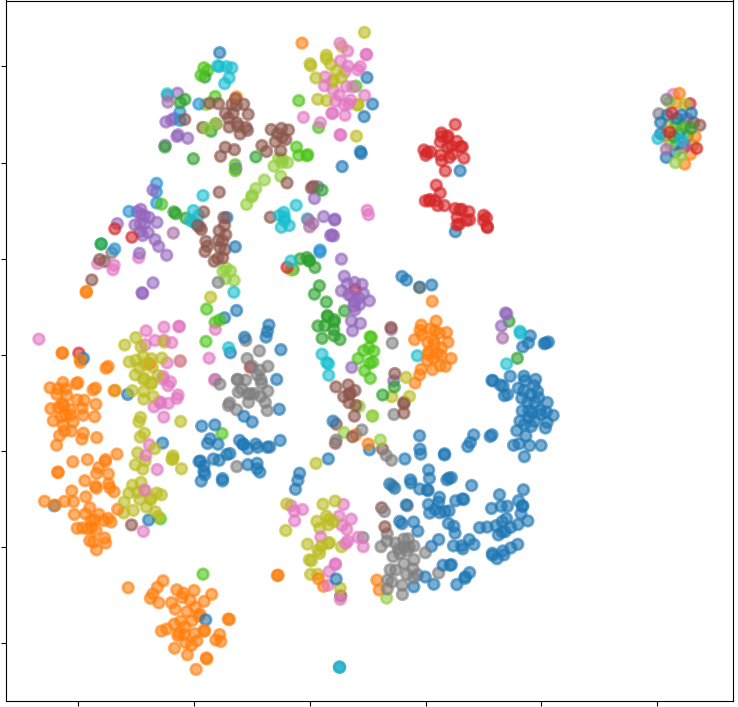}
    }
    \subfigure[$MGTCOM$ + $L_{\mathcal{T}}$]{
    \includegraphics[width=0.23\textwidth]{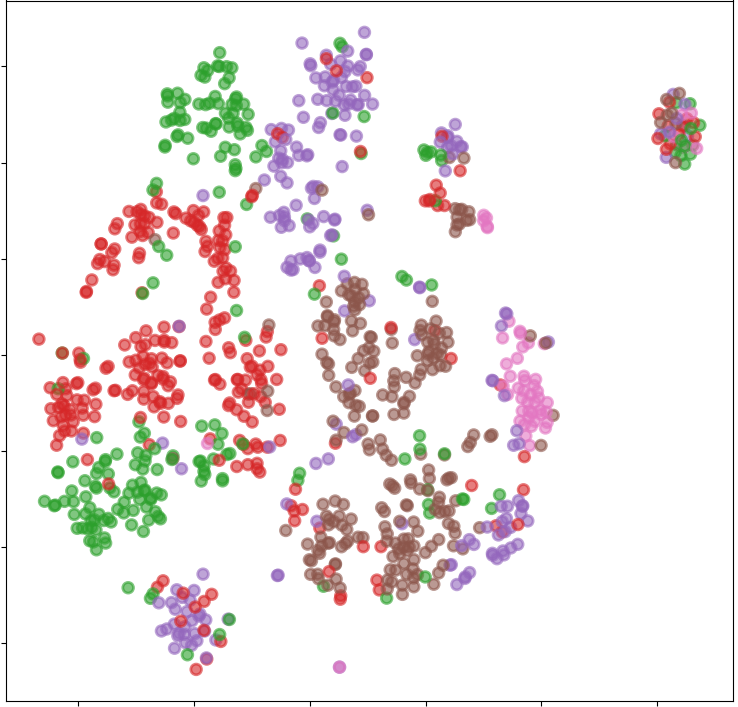}
    }
    \subfigure[$ComE$ + $L_y$]{
    \includegraphics[width=0.23\textwidth]{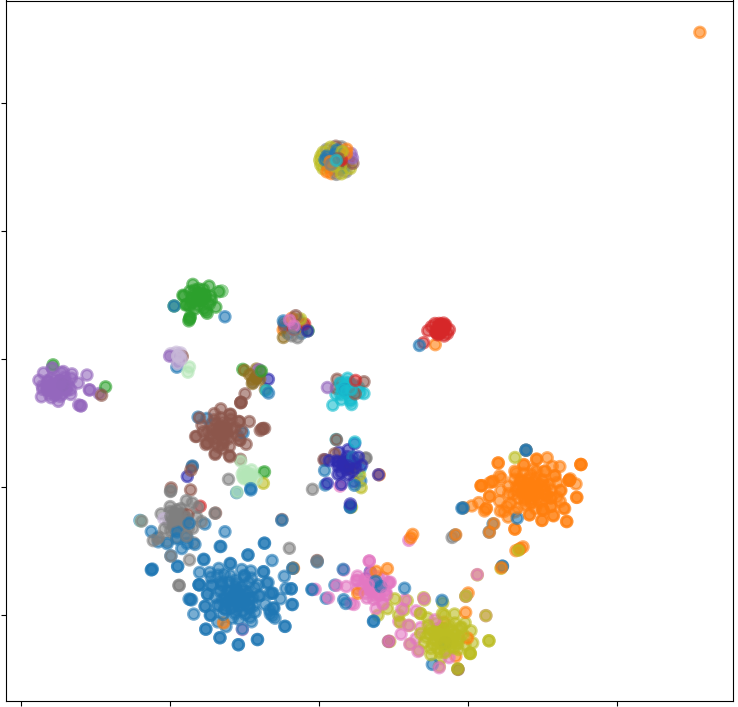}
    }
    \subfigure[$ComE$ + $L_{\mathcal{T}}$]{
    \includegraphics[width=0.23\textwidth]{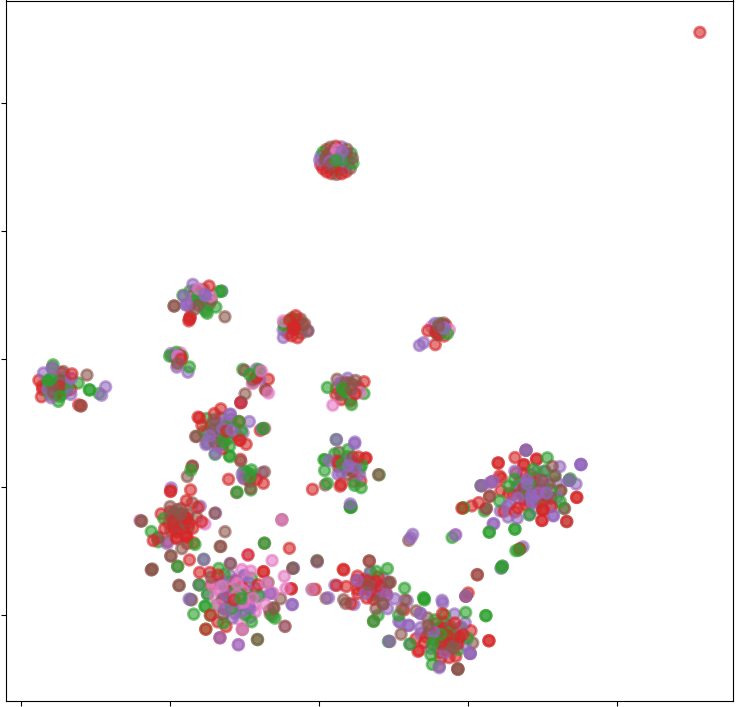}
    }\\
    
    \subfigure[$Node2Vec$ + $L_y$]{
    \includegraphics[width=0.23\textwidth]{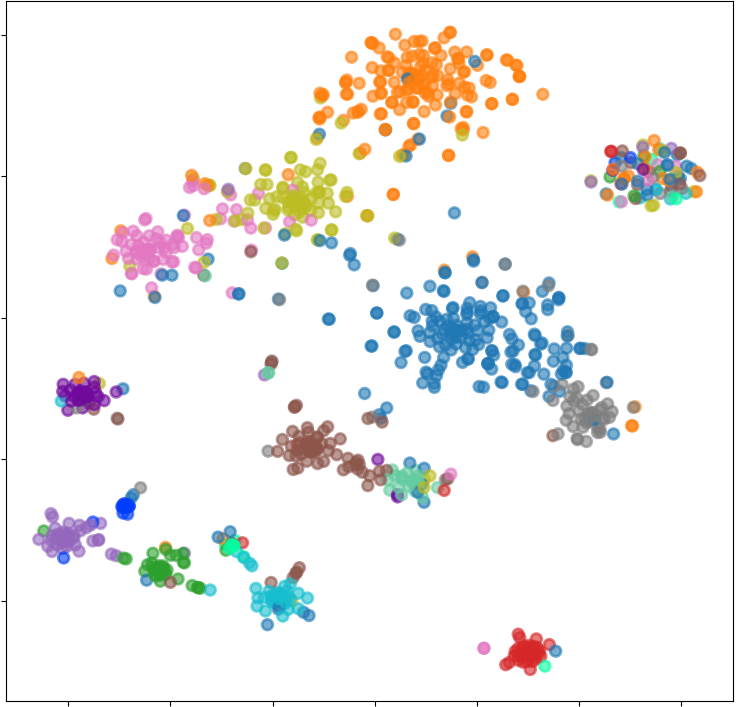}
    }
    \subfigure[$Node2Vec$ + $L_{\mathcal{T}}$]{
    \includegraphics[width=0.23\textwidth]{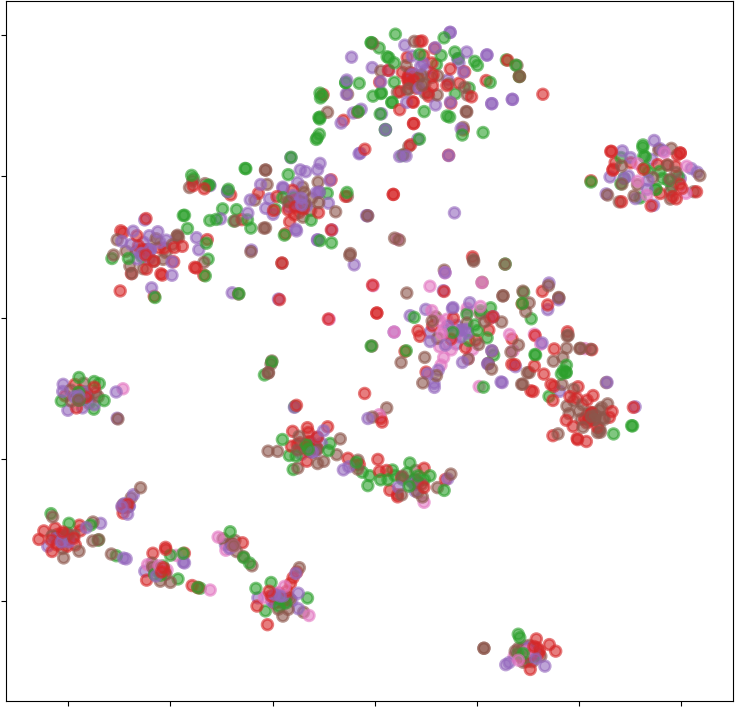}
    }
    \subfigure[$CTDNE$ + $L_y$]{
    \includegraphics[width=0.23\textwidth]{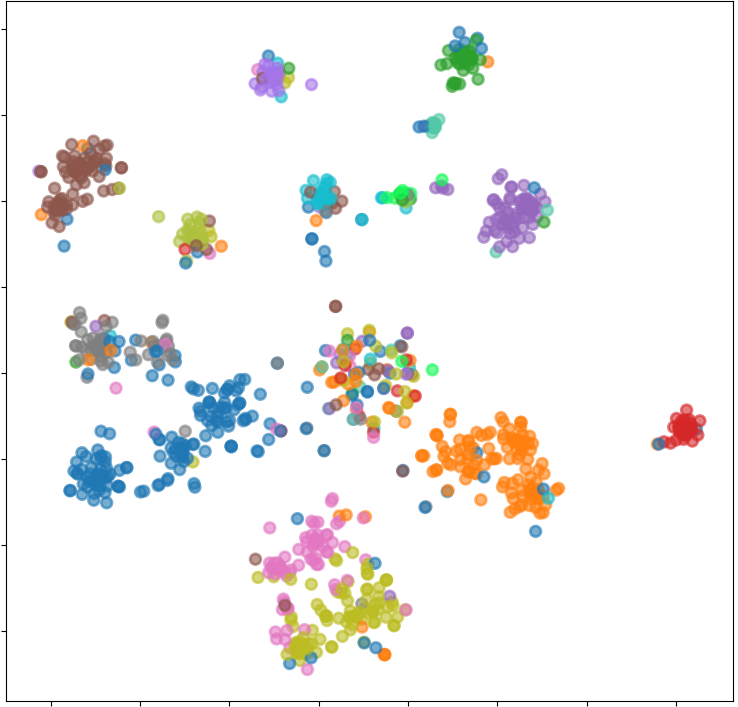} 
    }
    \subfigure[$CTDNE$ + $L_{\mathcal{T}}$]{
    \includegraphics[width=0.23\textwidth]{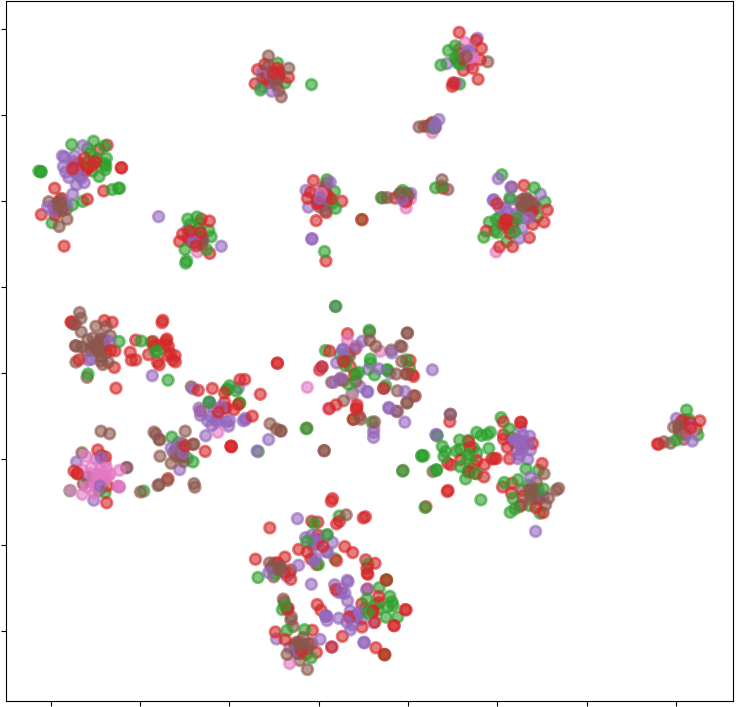}
    }
    \centering
    \small \textbf{(I)}
\end{minipage}%
\begin{minipage}{.33\textwidth}
    \includegraphics[width=\textwidth]{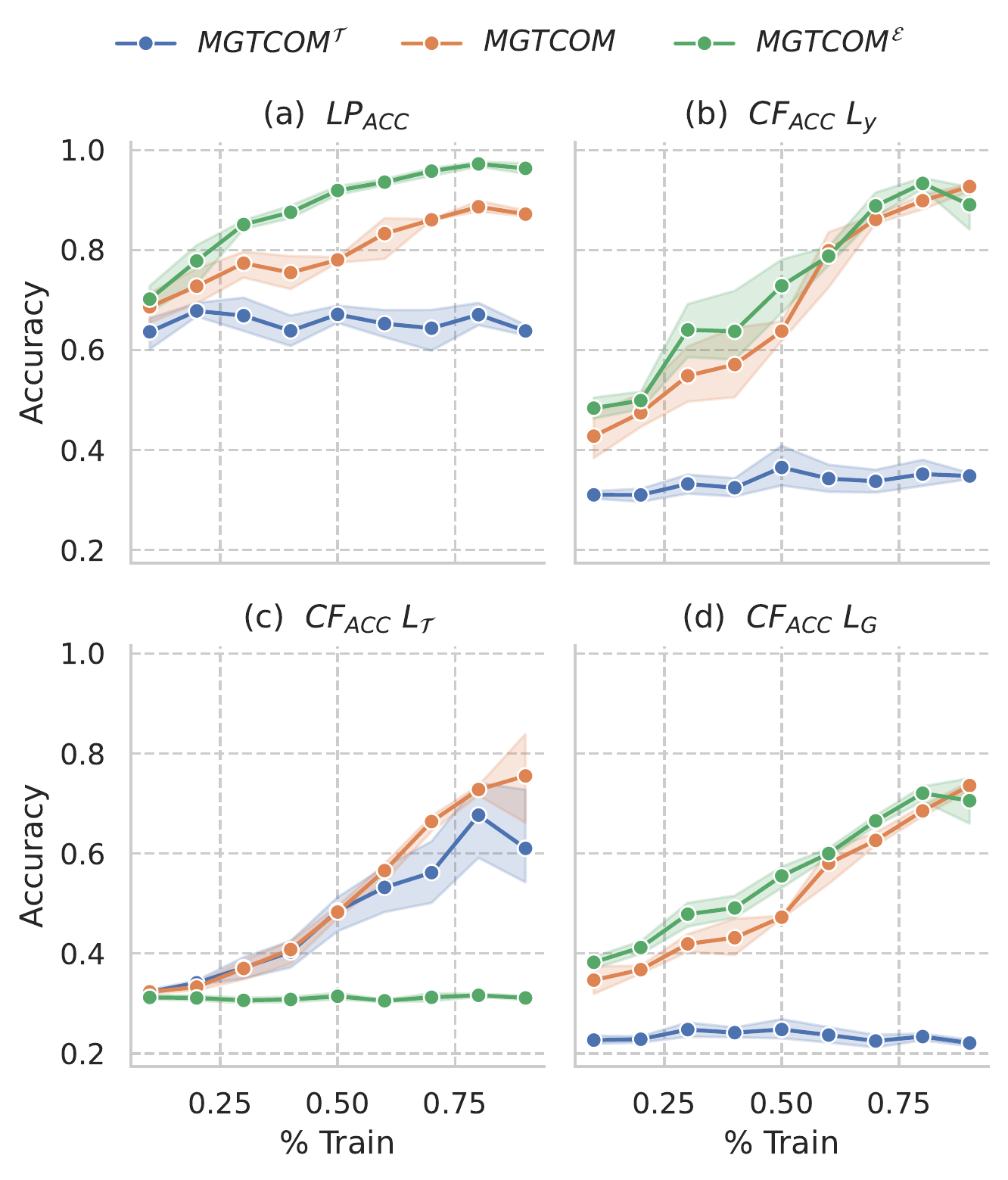}
    \centering
    \small \textbf{(II)}
\end{minipage}%


\caption{
(I) T-SNE plot of trained embeddings against ground truth ($L_y$) and timestamp labels ($L_{\mathcal{T}}$), and 
   (II) inductive inference results for  MGTCOM variants on the DBLP dataset.
}
\label{fig:combine}
\vspace{-2mm}
\end{figure*} 



\smallskip
\noindent
\textit{Hyperparameters}.
The hyperparameters for MGTCOM model can be attributed to either network architecture, topological random walk, temporal random walk or clustering.
The most important hyperparameters are specified below.
For primary embedding, we use two HGT layers with neighborhood sampling sizes of 8 and 4.
All the hidden dimensions are equal to the representation dimension, which is 64 ($d=64$).
For temporal and topological context sampling we use walk length $l=10$ (10 walks per node).
Node2Vec is configured to use $q=0.5$ to favor neighborhood exploration. 
The temporal sampling window $\omega$ for ballroom walk is determined for each dataset by splitting $\mathcal{T}$ into 20 even partitions.
For the clustering module we set prior parameters as $\nu = d + 1, \kappa = 1, \alpha = 10$ and $\Sigma_{scale} = 0.05$.
We set trade-off parameters as $\beta^{\mathcal{E}} = 1, \beta^{\mathcal{T}} = 1, \beta^{C} = 0.01$. 
For max-margin loss we set $\Delta$ to $0.1$.

\noindent
\textit{Baselines}. 
As baselines, we use various graph embedding and community detection algorithms, namely, ComE, GEMSEC, CP-GNN, CTDNE, GraphSAGE, Node2Vec and Matepath2Vec. We use the hyperparameters reported in their respective papers. 
To keep the results comparable, we use representation dimension $d=64$ for all.

\smallskip
\noindent
\textit{Datasets}. 
We use four real-world  multimodal datasets for evaluation, namely,
 DBLP~\cite{yangDefiningEvaluatingNetwork2012}, IMDB~\cite{IMDB5000Movie}, SDS~\cite{wangPublicSentimentGovernmental2020},  and ICEWS~\cite{garcia-duranLearningSequenceEncoders2018}. 
These datasets are of different types and contain information on various modalities.
We extracted a subset of the multimodal SDS graph from Twitter and applied additional preprocessing on the IMDB, DBLP, and ICEWS datasets to include the multimodal features present in the datasets but often not included in the graph due to sparsity of temporal or content-based features. 
Textual features are transformed using a pre-trained language model \cite{wangMiniLMDeepSelfAttention2020} and categorical features are encoded as one-hot feature vectors.
See \cref{tab:datasets} for a detailed comparison of node features.
For evaluation, we split the network edges into disjoint training (80\%), validation (10\%), and testing (10\%) sets.  
During link prediction, we exclusively use links in the respective set as positive pairs.
Negative pairs are sampled given the full set of edges.
Similarly, the clustering is computed on the training embeddings while cluster-based metrics are calculated using test and validation sets.
During the calculation of predictive metrics such as link prediction and classification, we run logistic regression three times and average the results. 

\subsection{Performance comparison}

We evaluate the performance of learned node embeddings and detected clusters.
In particular, we evaluate the predictive quality of embeddings using classification and link prediction, i.e., link prediction accuracy $LP_{ACC}$, temporal $L_\mathcal{T}$ and ground truth $L_y$  label classification accuracy $CF_{ACC}$.
We evaluate the quality of detected clusters by calculating their NMI score based on predefined ground-truth communities $L_y$, $L_\mathcal{T}$, $L_G$.
This tells us whether detected clusters approximate user-defined communities $L_y$, temporal partitioning $L_\mathcal{T}$ or the topology $L_G$.
Additionally, we calculate community quality scores for the learned community assignments using the modularity metric.

The embeddings obtained from non-community detection methods (GraphSAGE, Node2Vec and Metapath2Vec) were clustered using k-means clustering with $K=20$. Similarly, we use $K=20$ for community detection methods (ComE, GEMSEC, CP-GNN) that assume a predefined cluster count.
The results are reported in \cref{tab:results_perf}. 
It can be seen that while MGTCOM is competitive on task-specific measures such as link prediction and timestamp prediction, the community detection methods still have an edge on link-based modularity measures. 
A possible explanation for this would be the fact that the DPMM process is more prone to getting stuck in local minima as the clusters split and merge.
Another possibility is that node features do not contain enough information to model very specific network features such as modularity.

While CTDNE performs comparatively well in capturing the temporal aspect of a given network, we see that it still yields inferior results on datasets where temporal features are weakly correlated with topology.
It is interesting to note that algorithms that rely on pairwise loss measures such as GraphSAGE and CP-GNN perform relatively well on classification-based measures while performing very poorly on community quality measures modularity.
A possible explanation for such observation is that the combination of neighborhood sampling and pairwise loss reinforces structural similarity despite having a large receptive field.
Our method successfully overcomes this issue by modifying Hinge loss to work in a context path setting (see \cref{sec:obj_fn}).
We also observe that the MGTCOM model performs well on both topology and temporal prediction tasks in comparison to its task-specific counterparts. Specifically, MGTCOM outperforms all the models, in all the metrics but modularity, on the multimodal dataset SDS. 
Note that the modularity metric of MGTCOM is inconclusive because it mainly measures the quality of topological communities by looking at homophily. Heterophily commonly found in multimodal networks is penalized. 
Additionally, it outperforms all the methods, on the temporal task $L_t$, in all of the datasets.

\subsection{Qualitative results}
\label{sec:qualitative}
In this experiment, we further compare MGTCOM and the baseline models on the DBLP network.
We apply the T-SNE dimensionality reduction technique to visualize the trained node embeddings in 2D space colored by the ground truth label and the node timestamp (see \cref{fig:combine}-(\MakeUppercase{\romannumeral 1})). 
Since in this dataset the timestamps are weakly correlated with its topology, we can see that topology embedding (and community detection) methods such as ComE and Node2Vec do not capture temporal relations of nodes. On contrary, we observe distinct patterns emerge when looking at MGTCOM generated embeddings for both of the labels.
Similar to that of ComE the community structures are visible in the node embeddings though they are not as distinct. Note that, in \cref{fig:combine}-(\MakeUppercase{\romannumeral 1}), the embeddings are calculated on the training dataset. Each of the plots contains a blob of nodes that have no edges in the training set due to validation split. None of the methods is equipped to handle disconnected nodes.

\subsection{Inductive inference results}
\label{sec:ind_inference}
Since the MGTCOM model operates on sampled neighborhood subgraphs, in contrast to other methods, it can operate in an inductive setting.
Meaning that it is not necessary to retrain the model to infer representation vectors for previously unseen nodes.
On the DBLP dataset, we evaluate the performance of MGTCOM and its task-specific variants in an inductive setting by controlling the ratio of nodes in the training set to the validation set.
The test set remains constant throughout the experiment to accurately assess performance on inferred nodes.
%
In \cref{fig:combine}-(\MakeUppercase{\romannumeral 2}) we see the same measures ($LP_{ACC}$ and $CF_{ACC}$)  plotted with the training ratio on the x-axis.
From \cref{fig:combine}-(\MakeUppercase{\romannumeral 2}) (a) we observe that varying training set size does not affect link-prediction tasks as much as node classification tasks (b, c, d).
In all the measures, we can see that using only 75\% of the data does not substantially affect the results.
Interestingly, we observe that the variance on the temporal prediction task increases when more data is provided.

\section{Conclusion}
In this paper, we introduce the MGTCOM framework for community detection in multimodal graphs.  It utilizes meta-topological, topological, content features, and temporal information to detect communities.
Moreover, we address common issues in multimodal graphs such as information incompleteness, and inference on unseen data by adopting a graph convolutional network architecture that combines h-hop neighborhood sampling and random walk context sampling.
We devise a unified objective and an efficient temporal sampling method to learn multimodal community-aware node embeddings in an unsupervised manner.
Consequently, we leverage a split/merge-based Dirichlet process mixture model for community detection where the number of communities is not known a priori.
Our empirical evaluation shows that MGTCOM outperforms state-of-the-art on the largest multimodal dataset with incomplete features (SDS). It also achieves the best results on temporal tasks.     




\bibliographystyle{ACM-Reference-Format}
\bibliography{refs}

\end{document}